\documentclass[preprint]{elsarticle}

\usepackage{amssymb}

\usepackage{amsmath}

\usepackage{color}
\newcommand{\MARKI}[1]{#1}
\newcommand{\MARKII}[1]{#1}

\journal{Planetary and Space Science}

\begin{document}

\begin{frontmatter}



\title{Fast E-sail Uranus entry probe mission}


\author[FMI]{Pekka Janhunen\corref{cor1}\fnref{Tartu}}
\ead{pekka.janhunen@fmi.fi}
\ead[url]{http://www.electric-sailing.fi}
\author[orleans]{Jean-Pierre Lebreton\fnref{Meudon}}
\author[FMI]{Sini Merikallio}
\author[FMI]{Mark Paton}
\author[pisa]{Giovanni Mengali}
\author[pisa]{Alessandro A.~Quarta}
\fntext[Tartu]{Also at Tartu University, Estonia}
\fntext[Meudon]{Also at Paris Observatory at Meudon, Paris, France}

\address[FMI]{Finnish Meteorological Institute, Helsinki, Finland}
\address[orleans]{CNRS-Orleans, Orleans, France}
\address[pisa]{University of Pisa, Italy}
\cortext[cor1]{Corresponding author}

\begin{abstract}
The electric solar wind sail is a novel propellantless space
propulsion concept. According to numerical estimates, the electric
solar wind sail can produce a large total impulse per propulsion
system mass. Here we consider using a 0.5 N electric solar wind sail
for boosting a 550 kg spacecraft to Uranus in less than 6 years. The
spacecraft is a stack consisting of the electric solar wind sail
module which is jettisoned roughly at Saturn distance, a carrier
module and a probe for Uranus atmospheric entry.  The carrier module
has a chemical propulsion ability for orbital corrections and it uses
its antenna for picking up the probe's data transmission and later
relaying it to Earth. The scientific output of the mission is similar
to what the Galileo Probe did at Jupiter. Measurements of the chemical
and isotope composition of the Uranian atmosphere can give key
constraints to different formation theories of the Solar System. A
similar method could also be applied to other giant planets and Titan
by using a fleet of more or less identical probes.
\end{abstract}


\end{frontmatter}


\section{Introduction}

The electric solar wind sail (E-sail) is a newly discovered concept of
propelling an interplanetary spacecraft by employing the thrust
produced by the natural solar wind plasma stream
\cite{paper1,RSIpaper}. According to numerical performance estimates
\cite{paper7a}, the E-sail has a high ratio of produced total impulse
per propulsion system mass. The E-sail can be used for many solar
system propulsive tasks, including inner planets
\cite{MengaliEtAl2008}, asteroid rendezvous and sample return missions
\cite{PHA,KY26}, asteroid deflection \cite{MerikallioAndJanhunen2010},
various continuous thrust non-Keplerian orbits \cite{NKO} and flyby
or orbiter missions to outer planets
\cite{QuartaAndMengali2010,QuartaEtAl2011}.

The subject of the paper is a preliminary analysis on how to deliver
an atmospheric probe to Uranus reasonably fast (less than 6 years) and
at low cost in comparison to traditional outer planet missions. The
main objective is to make the case for conducting NASA's Galileo probe type
measurements at Uranus, i.e.~to measure the chemical and isotopic
composition of the atmosphere. A second goal of the paper is to
describe, although with less depth than for Uranus, the case for
sending a similar probe using the same novel propulsion technique to
all giant planets and possibly Titan.

\section{Why an atmospheric probe to Uranus?}

Measuring the chemical and isotope composition of giant planet
atmospheres is important because such measurements can constrain
models of the early history of the Solar System
\cite{SpilkerEtAl2009}. Thus far, only the Jovian atmosphere has been
directly probed {\em in situ}. Some presently favoured models of Solar
System history such as the Nice model \cite{TsiganisEtAl2005} and its
later variants \cite{LevisonEtAl2011} predict that the giant planets
formed originally at different solar distances than where they are
located nowadays. Because the temperature of the protoplanetary disk
depended on the solar distance, the models give predictions of the
chemical and isotope composition of the giant planet
atmospheres. However, many of these predictions can only be checked
against measurements by {\em in situ} probing of the planetary atmospheres.

An early analysis of Uranus and Neptune entry probes was made
relatively soon after the NASA's Voyager flybys \cite{TauberEtAl1994}. More
recently, a planetary entry probe engineering study for Venus, Saturn,
Uranus and Neptune was carried out at ESA's Concurrent Design Facility
\cite{ESA-PEP-study}.

All giant planet atmospheres should be eventually probed and the most
effective way of doing so might be by using a fleet of identical or
rather identical E-sail equipped probes, each one targeted to its own
planet. However, in this paper we concentrate on Uranus. Once a Uranus
mission has been designed, the other cases can be obtained as follows:
\begin{itemize}
\item Jupiter has already been probed by NASA's Galileo spacecraft. If a fleet mission to all
giant planets is made, it might be worthwhile to measure Jupiter again
with instruments identical to those used on the other planets.
\item Except for somewhat higher entry speed which requires somewhat
heavier heat shield, Saturn mission is technically easier than Uranus
because of shorter traveltime and shorter telemetry distance.
If an E-sail based mission to Uranus proves to be feasible, a Saturn mission
would be expected to be feasible, too.
\item If a fleet mission to all giant planets is conducted, it might be a
  good idea to also add Titan to the set of targets. Similar
  considerations to Saturn apply to Titan except that the heat shield
  requirements are smaller because of lower entry speed due to less
  massive target.
\item A Neptune mission would be, in principle, similar to Uranus mission except for
  longer traveltime and longer telemetry distance. However, in the
  mission architecture that we will analyse in this  paper, the telemetry distance is
  not a major cost factor because the total probe data volume is modest and
  because the data can be downlinked to Earth by the carrier module
  at a low bitrate.
\end{itemize}
Hence, by analysing a Uranus entry probe mission explicitly we can
take the first step not only for designing a Uranus probe mission, but
also assessing the mission requirements for all the giant planets.

We do not consider orbiter missions in this paper because regardless
of the employed propulsion technology, orbiter missions with their more
complex and comprehensive scientific payloads typically fall into a
significantly higher cost category than atmospheric probe
missions. The other reason is that a fast E-sail trajectory may be
relatively speaking less feasible for an orbiter mission, because in
case of Uranus and Neptune, one would need a significant amount of
chemical propellant for the planetary orbit insertion or alternatively
one should rely on aerocapture whose technical readiness level is
lower than that of chemical propulsion.

\section{E-sail Uranus entry probe mission}

Our proposed E-sail Uranus entry probe mission consists of three modules
which are initially stacked together: the E-sail module, the carrier
module and the entry module. The entry module is composed of the
atmospheric probe inside a heatshield. The stack is initially launched to
Earth escape orbit by a conventional booster. The E-sail module
accelerates the stack to Uranus intercepting trajectory.  The carrier
module performs the necessary orbital corrections to fly by Uranus and
to collect the data transmitted by the atmospheric probe. In more detail, the mission proceeds according
to the following steps:
\begin{enumerate}
\item The stack is launched to Earth escape orbit by a conventional
  booster. Any escape orbit \MARKI{(i.e. any orbit with non-negative
    specific energy parameter C$_3$)} is suitable for the
  purpose. \MARKI{For the E-sail to work, it is required to be in the
    solar wind.}
\item The E-sail module accelerates the stack to a trajectory towards Uranus.
\item The E-sail module is abandoned approximately at Saturn distance.
\item The carrier module uses chemical propulsion \MARKII{(in this paper baselined as
  green monopropellant)} for orbital corrections.
\item About 13 million km (8 days) before Uranus, the carrier module
  detaches itself from the entry module and makes a $\sim 0.15$ km/s
  transverse burn so that it passes by the planet at $\sim 10^5$ km
  distance, safely outside the ring system. Also a slowing down burn
  of the carrier module may be needed to optimise the link geometry
  during flyby.
\item Protected by the heat shield, the entry module enters into atmosphere.
\item A parachute is deployed and the heat shield is separated.
\item The probe falls under parachute in Uranus atmosphere, makes scientific measurements
  and transmits data to the high gain antenna of the carrier which
  flies by at $\sim 10^5$ km distance.
\item After exiting Uranus environment, the carrier redirects its high gain antenna towards
  Earth to transmit the stored probe science data.
\end{enumerate}

The communication frequency between probe and carrier cannot be set
too high because of the attenuation and scattering of the radio signal
caused by atmospheric gases and clouds. The frequency selection
trade-off study is outside the scope of this paper, but using the same
values as Galileo probe should be a good starting point. The Galileo probe
worked until 20 bar pressure and used 1.39 GHz frequency to
transmit 3.5 Mbit of data volume to the orbiter's 1.1 m
receiving antenna \cite{GalileoReview1992}. The mass of the Galileo
orbiter hardware dedicated to receiving and relaying the
probe data was 23 kg \cite{GalileoReview1992}. In our case,
communication with Earth would probably be done at higher frequency
although using the same $\sim 1$ m antenna dish. A one metre dish
enables only slow communication with Earth from Uranus distance, but
this is not a problem since the carrier spacecraft has plenty of time
to send the data after passing by the planet.

Table \ref{table:toplevel} shows the top-level mass budget. The 0.5 N
E-sail module component masses (Table \ref{table:Esailmodule}) are
adopted from the last column of Table 3 of Ref.~\cite{paper7a}. \MARKII{The design
uses 50 tethers, each of which is 18 km long, and made
of 50 $\mu$m diameter aluminium base wire and three 25 $\mu$m loop
wires whose purpose is to prevent the tether from breaking even when
micrometeoroids randomly cut its individual wires \cite{SeppanenEtAl2013}.} We assume
that the auxiliary tethers are made of 7.6 $\mu$m thickness kapton
which is currently an ITAR-restricted product. \MARKII{The purpose of
  the auxiliary tethers is to connect together the tips of the main tethers 
  to ensure dynamical stability of propulsive flight \cite{RSIpaper}.} Using ITAR-free 12.6
$\mu$m kapton would increase the mass of the E-sail module by 12.1
kg. Regardless of the used thickness, it could be possible to save
some mass by using a more aggressive punching pattern for the
auxiliary tethers. We included a 20\% E-sail module system margin.

\begin{table}[t]
\caption{Top-level mass budget.}
\centering
\begin{tabular}{ll}
\hline\noalign{\smallskip}
E-sail module & 150 kg \\
Carrier module (wet) & 150 kg \\
Entry module & 256 kg \\
\hline\noalign{\smallskip}
Total & 556 kg \\
\hline
\end{tabular}
\label{table:toplevel}
\end{table}

\begin{table}[t]
\caption{E-sail module key properties and mass budget.}
\centering
\begin{tabular}{ll}
\hline\noalign{\smallskip}
0.5 N thrust at 1 au from the Sun & \\
0.9 mm/s$^2$ characteristic acceleration & \\
\hline\noalign{\smallskip}
50$\times$18 km main tethers ($50\mu$m base$+3\times 25 \mu$m loop wire) & 10.3 kg \\
Main tether reels & 11.4 kg \\
Electron guns & 1.59 kg \\
540 W/40 kV high voltage source & 10.6 kg \\
Tether cameras and E-sail control electronics & 1.48 kg \\
50 Remote Units & 49.3 kg \\
7.6 $\mu$m$\times 3$ cm 50\% punched kapton auxiliary tether ring & 18.2 kg \\
E-sail module structural & 22 kg \\
E-sail system margin +20\% & 25 kg \\
\hline\noalign{\smallskip}
E-sail module total & 150 kg \\
\hline
\end{tabular}
\label{table:Esailmodule}
\end{table}

For passive dynamical stability in the E-sail deployment and cruise
phase, the inertial moment of the spacecraft stack should be largest
along the spin axis. In other words, the spacecraft stack should
resemble a disk or relatively flat cylinder which has the tethers attached along its
perimeter. Each tether attachment point must also have room for storing
the corresponding Remote Unit before deployment. If 25 cm is enough
for each stowed Remote Unit, then the length of the perimeter must be
$50 \times 0.25$ m $= 12.5$ m, corresponding to 4 m diameter
disk. This fits into launchers such as Soyuz, although it exceeds the
diameter of the payload fairing of small launch vehicles.
If compatiblity with small launchers is desired, the tether attachment
ring must be deployed from a more compact configuration. We have done
some in-house (Finnish Meteorological Institute) prototyping work on how such tether ring deployment
could be done and the initial results look promising.

Alternatively, one could reduce the spacecraft diameter by stacking
the tethers in more than one vertical layer. For example, if two layers
are used then the disk diameter can be 2 m, a value compatible with
small launchers. In this case the spacecraft would look more like a cylinder than
a disk. It would still seem feasible to have all components of the
stack (E-sail, carrier and entry module) with flat enough shape such that
the maximum inertial moment occurs along the cylinder axis. If not, as
a fallback solution, the requirement of the inertial moment and passive
dynamical stability could be relaxed by resorting to active attitude
control during E-sail deployment and cruise phases.

At 1 au, the 0.5 N E-sail needs nominally 540 W of electric power to keep its
tethers charged \cite{paper7a}.  The electric power requirement of the E-sail
scales as $1/r^2$ i.e. in the same way as the illumination of solar
panels \cite{RSIpaper,ToivanenAndJanhunen2009} \MARKI{although the thrust
  scales as $1/r$ \cite{RSIpaper}}. Thus, if enough solar
panels are used to power the E-sail at 1 au, the same panel area is
sufficient also at larger solar distances, excluding a small constant
power needed by E-sail control systems. During cruise, the E-sail spin
plane is \MARKII{typically inlined by at most} $\sim 45^{\circ}$ with the \MARKII{solar} direction
so that the illumination of the solar panels is reduced by a factor of
$\sim 0.7$. \MARKII{This is a conservative estimate because significant inclination occurs early
  in the mission where solar illumination is strong.} If we require that the total power is 1 kW at 1 au (at 45$^{\rm
  o}$ orientation) and assuming 20\% overall efficiency for the
panels, then the required panel area is 5.35 m$^2$. This panel area
fits easily inside the 4 m diameter disk configuration which was
discussed above. It does not fit on a 2 m diameter disk area, however,
so that in the 2 m diameter cylinder solution which was discussed
above, one has to use deployable solar panels.

We consider the following principal options for the power system:
\begin{enumerate}
\item The solar panel power system is part of the E-sail module and
  is thus jettisoned with it. The carrier module is powered by a radioisotope
  thermoelectric generator (RTG) whose waste heat keeps the whole
  stack warm. After detachment, the entry module is powered by a
  \MARKI{primary} battery and kept warm by radioisotope heater units (RHUs).
\item The solar panel power system is part of the carrier module.
  The carrier module is powered by the low
  remaining solar panel power (2.5 W at Uranus distance) and by a
  \MARKI{primary} battery. All modules have RHUs for temperature management. The entry
  module is battery-powered as before. The carrier must have a very
  lower power hibernation mode. The benefit is that only RHUs, but no RTG are
  needed.
\end{enumerate}
\MARKI{The entry module needs $\sim 100$ W
  of power for maximum $\sim 2$ hours. The corresponding primary battery mass
  is $\sim 2$ kg. Low illumination low temperature (LILT) qualified
  solar panels must be used at Uranus distance or else one must use
  concentrators. Unconcentrated LILT triple junction cell
  efficiency of 25\% has been reported in tests mimicking Uranus
  distance \cite{PiszczorEtAl2008}.}

The main properties of the carrier module are given in Table
\ref{table:carrier}. The carrier module contains an attitude control
system, a power system (as discussed above), a high-gain parabolic
dish antenna of $\sim 1$ m diameter and a chemical propulsion system,
baselined to use green monopropellant \cite{LMP103S} with specific
impulse \MARKI{of 255 s}. The chemical propulsion system is needed for making orbital
corrections after the E-sail cruise phase and for boosting the carrier
sideways and slowing it down after releasing the entry module to pass by Uranus at proper
$\sim 10^5$ km distance. We reserve 0.22 km/s delta-v for orbital
corrections and after probe detachment 0.3 km/s is available in total
for the 0.15 km/s transverse boost and a slowing-down boost which
improves the probe to carrier link geometry. Notice that after
separating from the entry module, the spacecraft is much more
lightweight so that the propellant budget is relatively insensitive to the
amount of delta-v needed after probe separation. The mass
budget of the carrier is an estimate which is not yet based on an
accurate calculation. Nevertheless we think that since the
requirements of the carrier are relatively simple, the mass budget is
probably not unrealistically low.

\begin{table}[t]
\caption{Properties of carrier module.}
\centering
\begin{tabular}{ll}
\hline\noalign{\smallskip}
Wet mass & 150 kg \\
Dry mass & 100 kg \\
\MARKII{Green monopropellant $I_{\rm sp}$} & \MARKI{255 s} \\
Total propellant & 50 kg \\
Propellant to use before entry module release & 37 kg \\
Propellant to use after entry module release & 13 kg \\
Delta-v capacity for orbital corrections & 0.22 km/s \\
Delta-v capacity after entry module release & 0.3 km/s \\
High gain parabolic antenna & Diameter $\sim 1$ m \\
Attitude control system \\
Receiver to pick up probe's transmission \\
Transceiver for Earth communication \\
\hline\noalign{\smallskip}
\end{tabular}
\label{table:carrier}
\end{table}

The scientific and environmental requirements of the entry module are
similar to the Jupiter Galileo probe except that the entry speed is
smaller so that a lower heat shield mass is sufficient. The Galileo
probe total mass was 339 kg of which 45 \% (152 kg) was the heat
shield. In our case we assume 30\% heat shield mass fraction and 179
kg bare mass. 

\begin{table}[t]
\caption{Mass budget of entry module.}
\centering
\begin{tabular}{ll}
\hline\noalign{\smallskip}
Total mass & 256 kg \\
Heat shield & 77 kg (30\% of total) \\
Total without heat shield & 179 kg \\
Bus & 143 kg (80\% of 179 kg) \\
Science instruments & 36 kg (20\% of 179 kg) \\
\hline\noalign{\smallskip}
\end{tabular}
\label{table:entry}
\end{table}

\begin{figure}
\centerline{\includegraphics[width=0.75\columnwidth]{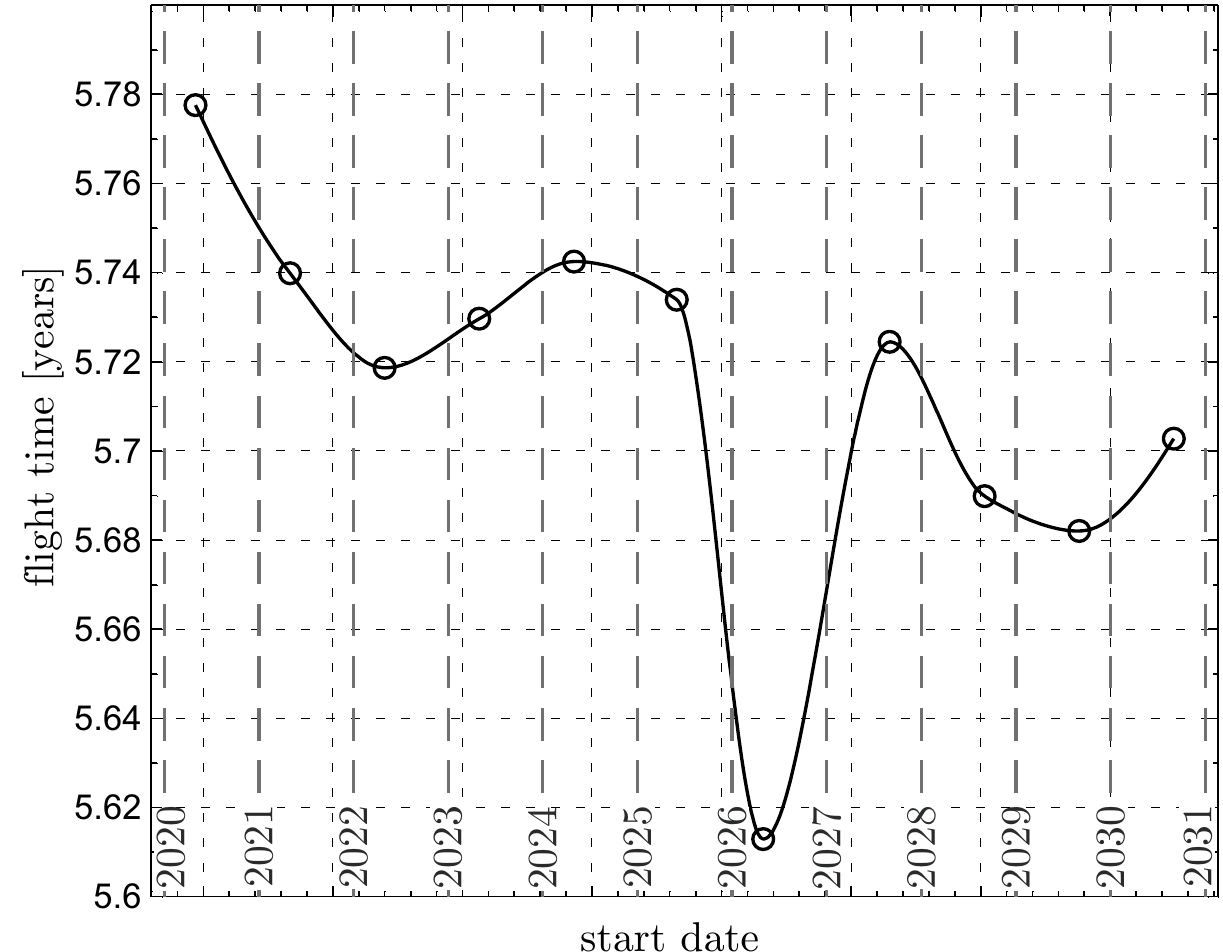}}
\centerline{\includegraphics[width=0.75\columnwidth]{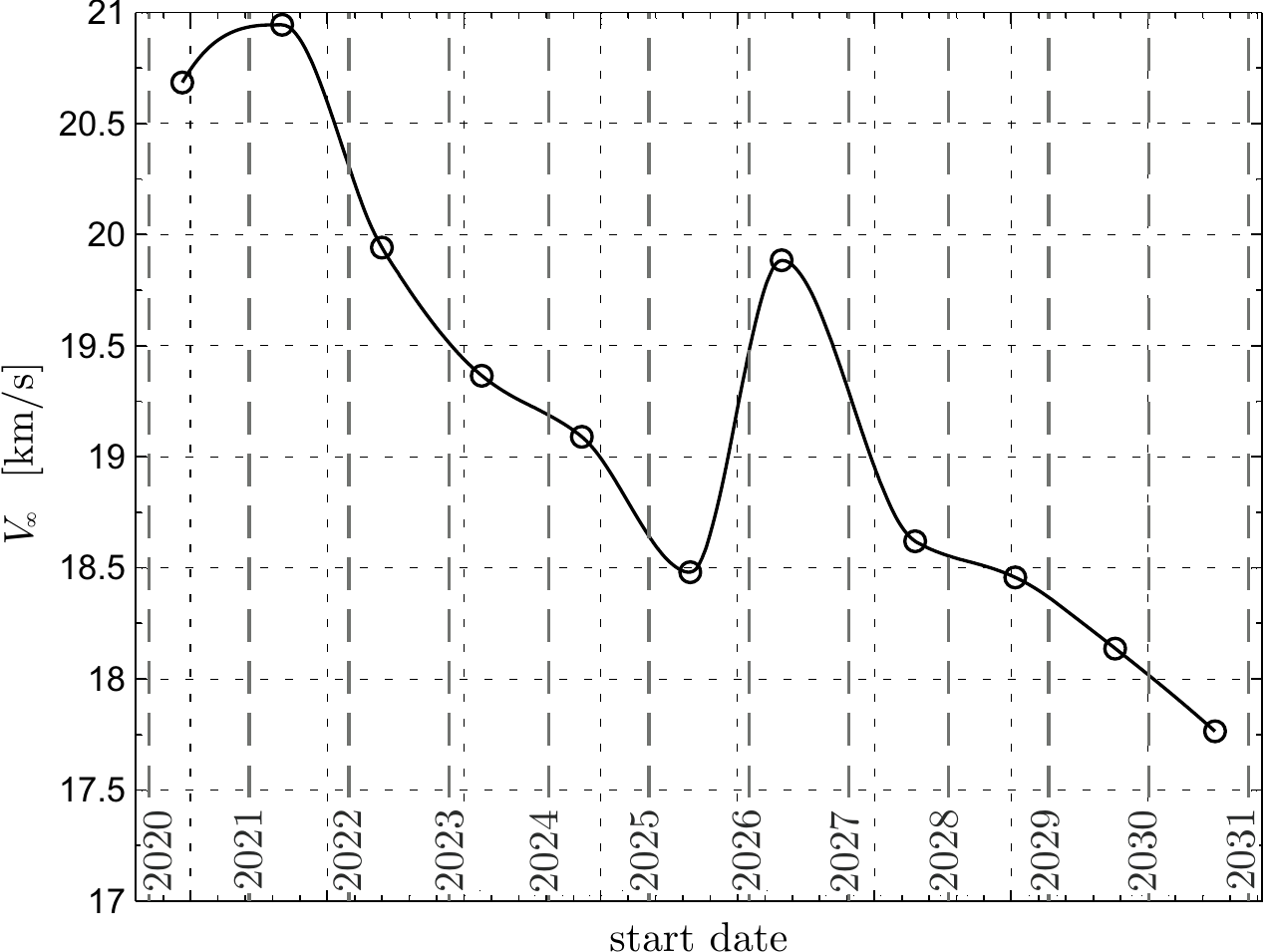}}
\caption{
Top: minimum traveltime from Earth to Uranus for launch year in
2020-2030. Bottom: the corresponding
hyperbolic excess speed $V_\infty$ at Uranus.
\label{fig:flighttimes}
}
\end{figure}

Figure \ref{fig:flighttimes} shows the traveltime from Earth to Uranus
\MARKII{(starting from Earth C$_3$=0)} using an E-sail spacecraft with 0.9 mm/s$^2$ characteristic
acceleration. The characteristic acceleration is the maximum
propulsive acceleration when the Sun-spacecraft distance is 1 au. An optimised low thrust orbit was computed using
real planetary emphemerides and using the first day of each month as a
starting date. Maximum usable thrust vector coning angle of 30$^{\rm
  o}$ was assumed and the E-sail was turned off at 9 au. For each
calendar year in 2020-2030, the minimum of the obtained 12 monthly
trajectory traveltimes was calculated and plotted in
Fig.~\ref{fig:flighttimes}. The corresponding Uranus-approaching
hyperbolic excess speed $V_\infty$ is also plotted in Fig.~\ref{fig:flighttimes}.
By $V_\infty$ we mean the relative speed of the probe with respect to
Uranus, computed outside the planet's gravity well but near the
planet from the heliocentric perspective (within the framework of the
patched conic approximation). As seen in Fig.~\ref{fig:flighttimes}, a
typical value is $V_\infty=20$ km/s.

\begin{figure}
\centerline{\includegraphics[width=0.75\columnwidth]{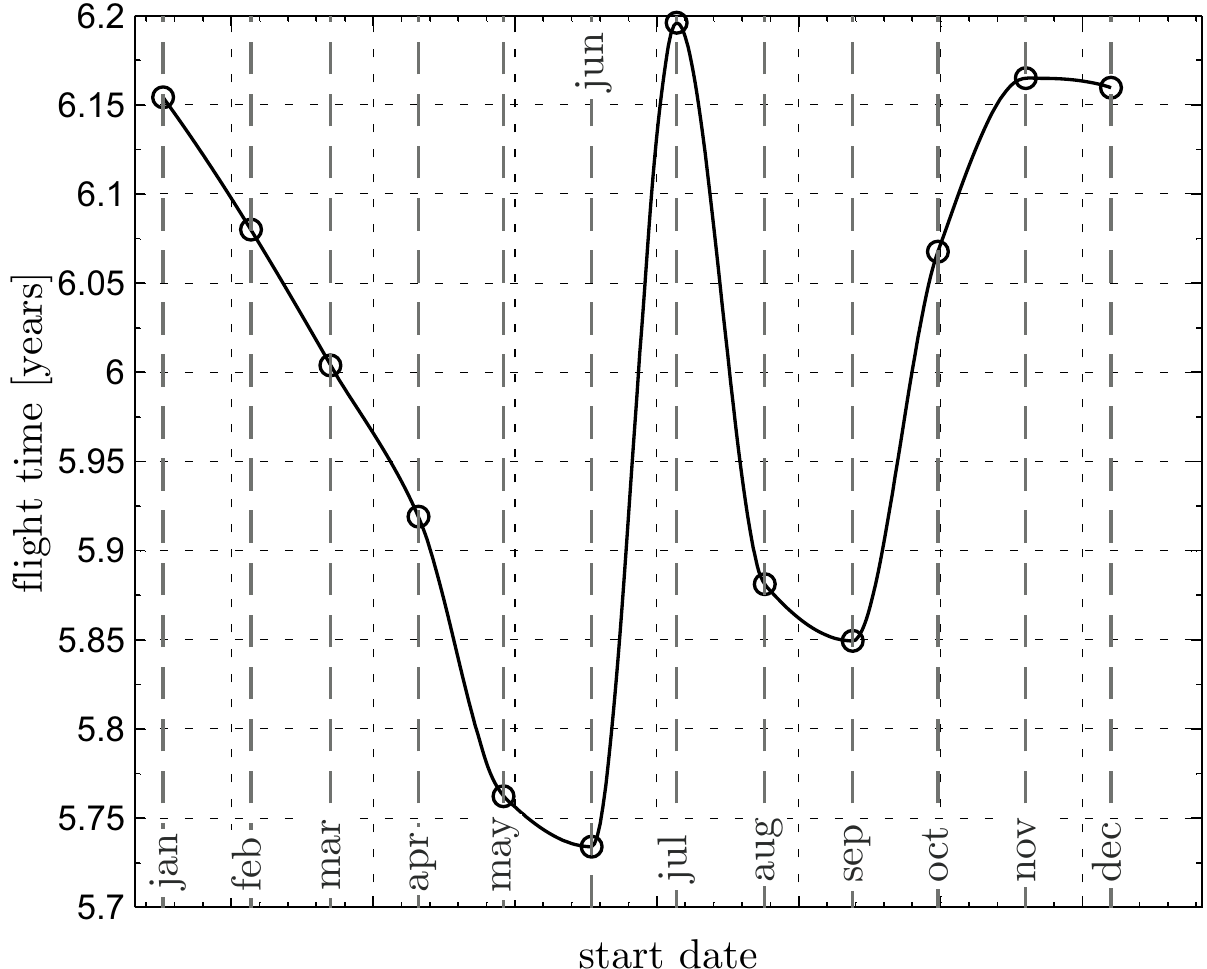}}
\caption{
Dependence of traveltime on starting month in exemplary year 2025.
\label{fig:flighttimes2025}
}
\end{figure}

In Figure \ref{fig:flighttimes2025} we show the dependence of the
traveltime on the starting month for an exemplary year of 2025. The
difference in traveltime between the optimal month and the least favourable month is
only 0.5 years. Figs.~\ref{fig:flighttimes} and
\ref{fig:flighttimes2025} show that in stark contrast to traditional
mission architectures relying on gravity assist manoeuvres, \MARKII{with}
the E-sail the \MARKII{dependence of the} traveltime on the starting
date \MARKII{is feeble enough} that one could in practice launch the probe at any time.

\begin{figure}
\centerline{\includegraphics[width=0.75\columnwidth]{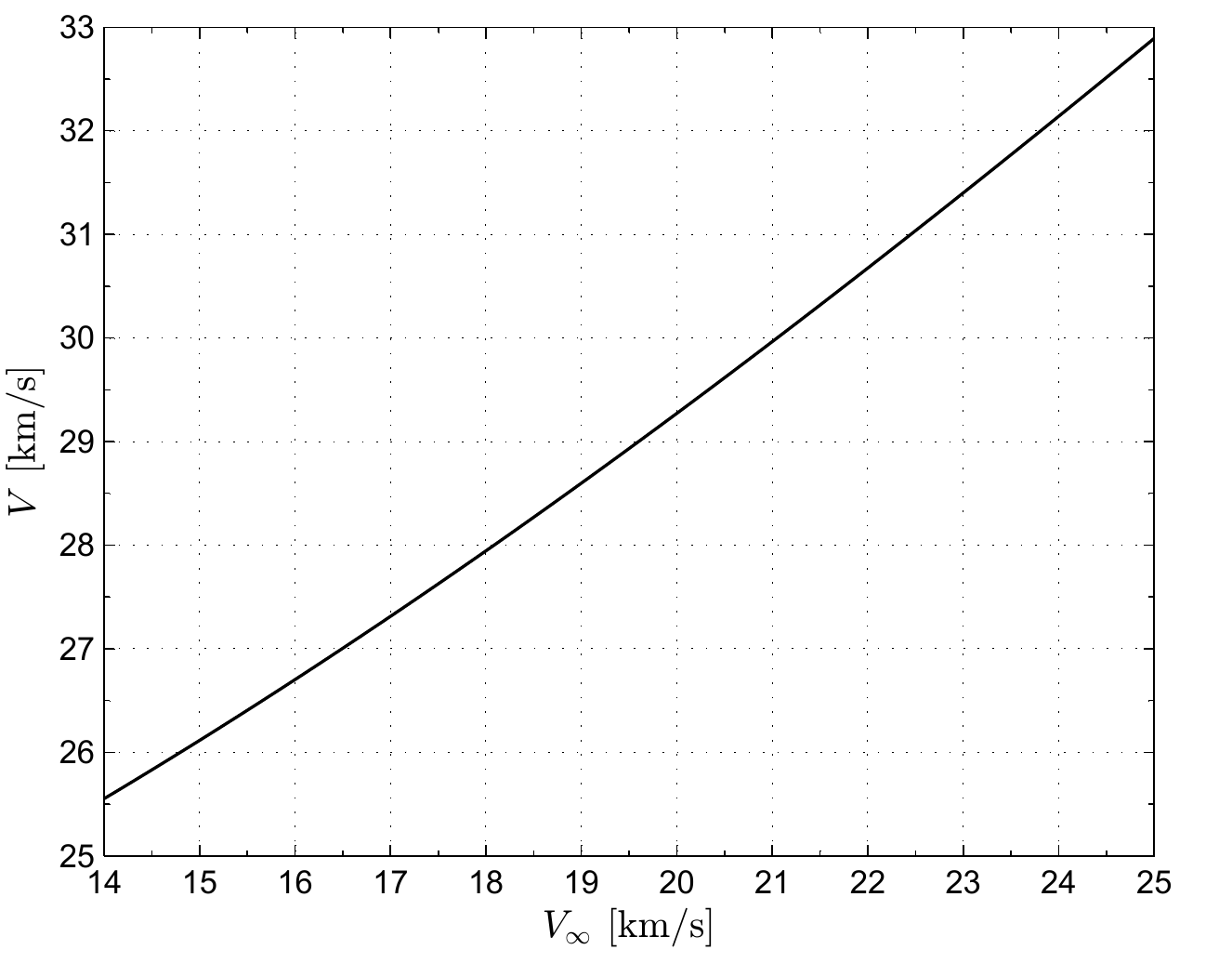}}
\caption{
Atmospheric entry speed $V$ (not taking into account planetary rotation)
as function of hyperbolic excess speed $V_\infty$.
\label{fig:entryspeed}
}
\end{figure}

Figure \ref{fig:entryspeed} shows the atmospheric entry speed $V$ as a
function of the hyperbolic excess speed $V_\infty$. For the typical
value of $V_\infty=20$ km/s the entry speed is $V\approx 30$ km/s.

\begin{figure}
\centerline{\includegraphics[viewport=150 400 450 750,width=0.7\columnwidth]{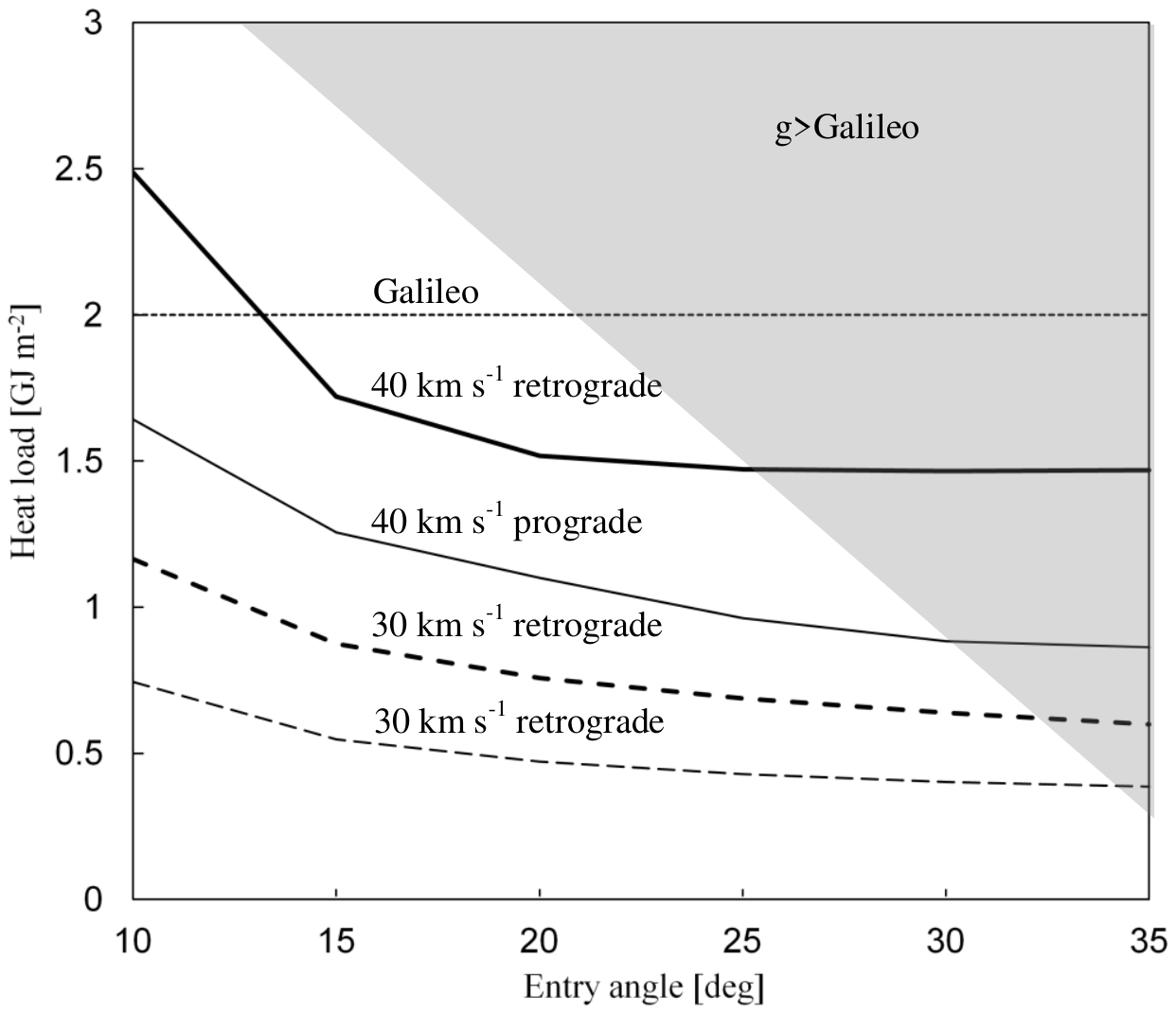}}
\caption{
Modelled atmospheric entry heat load encountered by the probe, as
function of entry angle and for different entry speeds $V$. The heat
load of the Galileo probe is marked. Within the shaded region, the maximum
deceleration exceeds the value encountered by the Galileo probe.
\label{fig:heatload}
}
\end{figure}

In Figure \ref{fig:heatload} we show the total heat load encountered
by the probe as function of the entry angle and for various entry
speeds $V$. The entry speed $V$ is computed with respect to
nonrotating planet; in Fig.~\ref{fig:heatload} cases of maximum (equatorial)
prograde and maximum retrograde entries are shown separately. For the
typical $V=30$ km/s entry speed, the heat load remains by factor 3-4
below the Galileo value for all values of the entry angle. If the
entry angle is steeper than 30$^{\circ}$-35$^{\circ}$, the maximum deceleration
experienced by the probe goes beyond the Galileo value, however. Based
on Fig.~\ref{fig:heatload} we estimate that 30\% heat shield mass
fraction (Table \ref{table:entry}) is \MARKII{conservative} and leaves
rather large freedom for the selection of the entry angle in the range
10$^{\circ}$-30$^{\circ}$.

\begin{figure}
\centerline{\includegraphics[width=0.75\columnwidth]{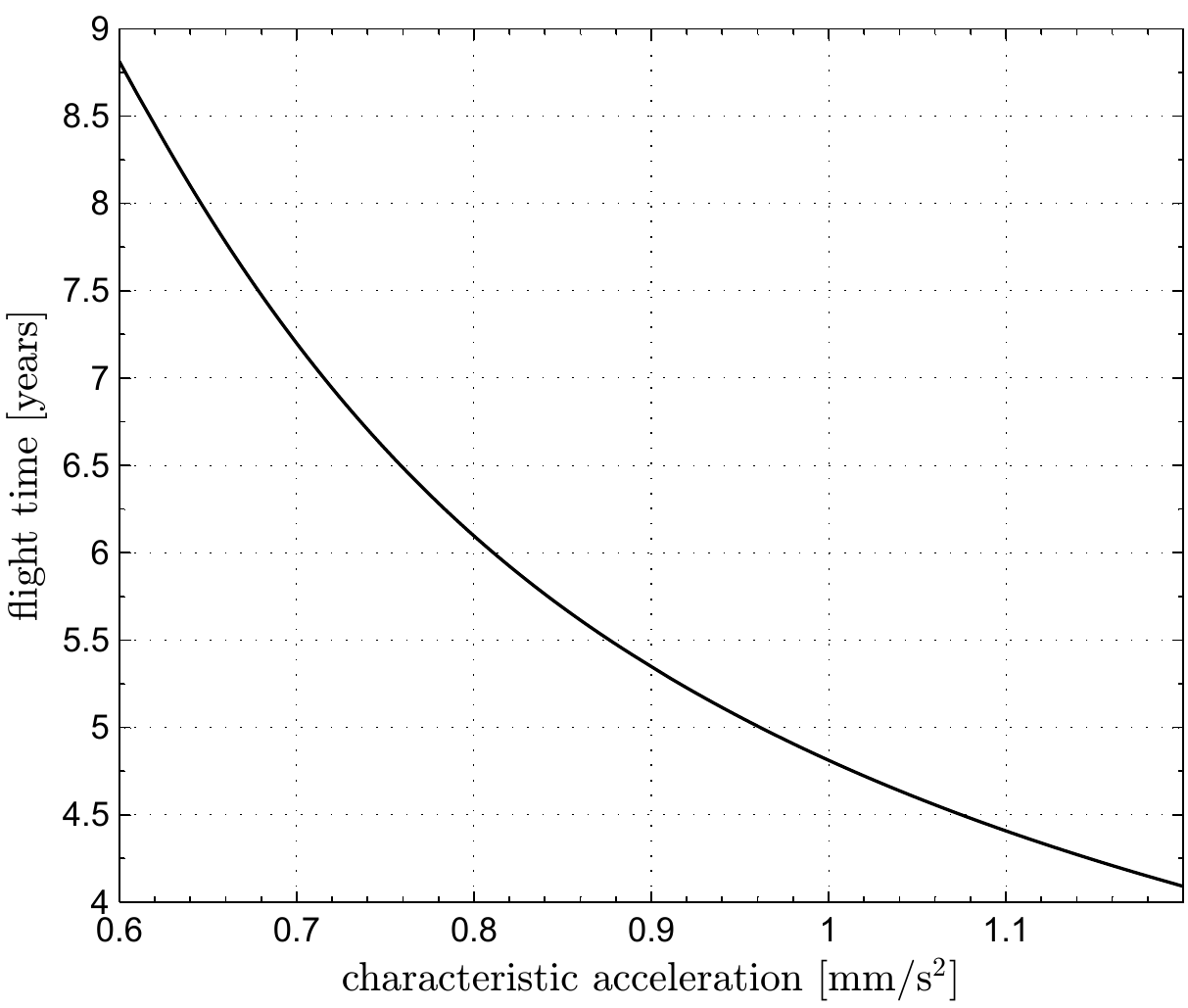}}
\centerline{\includegraphics[width=0.75\columnwidth]{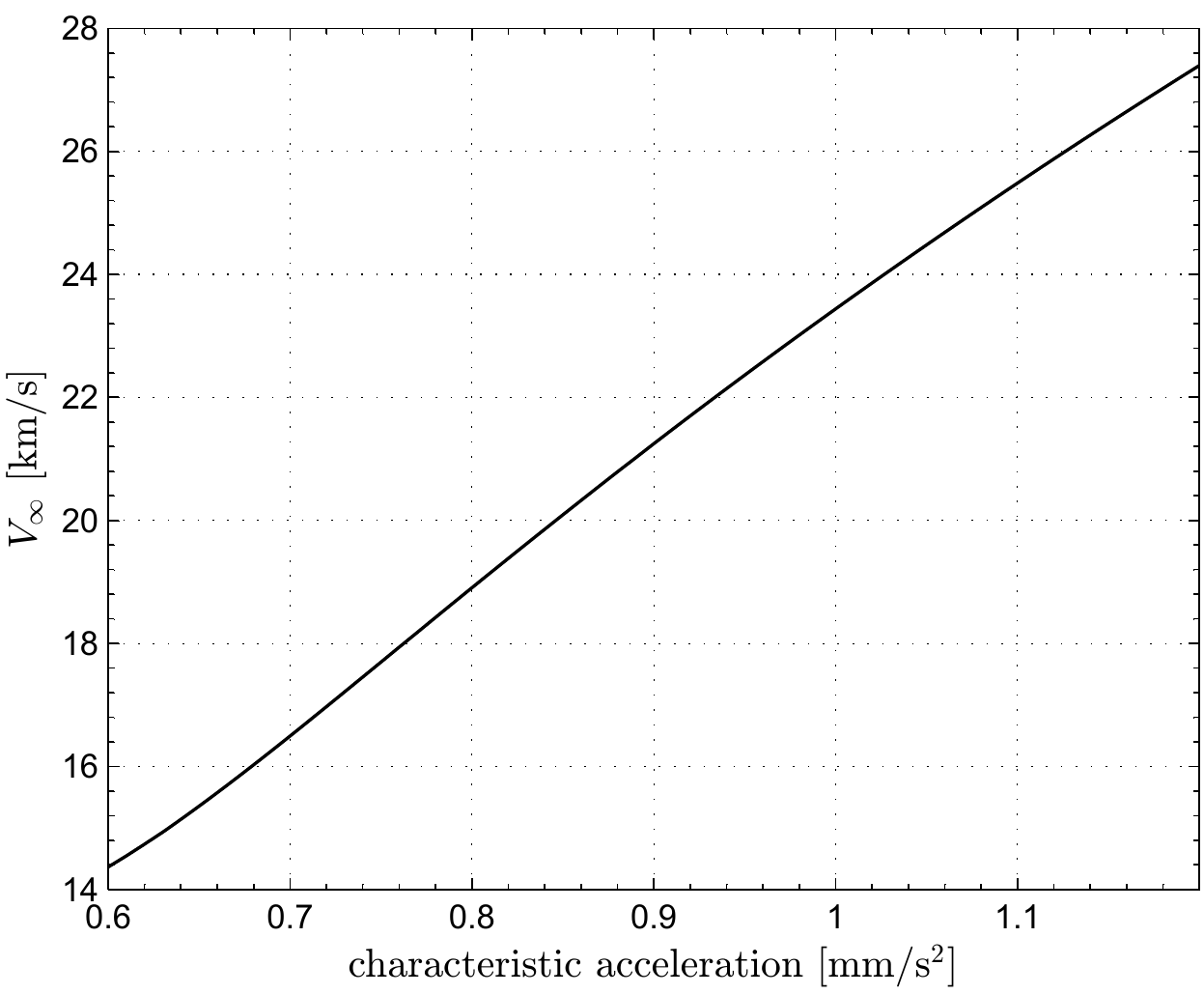}}
\caption{
\MARKI{
Dependence of traveltime (top) and Uranus atmospheric entry speed
(bottom) on characteristic acceleration whose baseline value is 0.9 mm/s$^2$.
\label{fig:flighttimedep}
}
}
\end{figure}

\MARKI{

\subsection{Analysis of sensitivity of traveltime to mass}

Figure \ref{fig:flighttimedep} shows the dependence of the traveltime
and Uranus atmosphere entry speed on the spacecraft characteristic
acceleration. To simplify calculations, we assumed circular coplanar
Earth and Uranus heliocentric orbits without ephemeris constraints in
Fig.~\ref{fig:flighttimedep}, while elsewhere in the paper we use full
3-D orbit calculations.

Let us consider an artificial case example where the wet mass of the carrier module is
increased by 20\%, from 150 kg to 180 kg. The total mass then
increases from 556 kg to 586 kg (5.4 \% increase).
The characteristic acceleration gets decreased from 0.90 to 0.854 mm/s$^2$,
and from Fig.~\ref{fig:flighttimedep} one can infer that the
traveltime increases by 7.5 \%, or by 5 months. Also one sees from
Fig.~\ref{fig:flighttimedep} that the entry speed
is lowered by 4.7\% from 21.2 to 20.2 km/s so that the 
dissipated energy is reduced by by 9\%. This enables one to reduce
the heat shield mass (originally 77 kg) by approximately a proportional amount (7 kg),
thus cancelling 23 \% of the originally assumed 30 kg mass increase.

}

\section{Discussion and conclusions}

We showed that given a working E-sail with currently projected
performance characteristics \cite{paper7a}, \MARKII{one could deliver
  an atmospheric probe mission to Uranus in less than six years. The
  mission would have Earth escape (C$_3$=0)} mass of 500-600 kg and the
launch \MARKII{would be} possible at any time without launch window
constraints. Such mission could greatly contribute to the
understanding of the history of the Solar System and especially
possible giant planet migration.

\MARKI{Obviously, before these benefits can be realised, an E-sail
  based technology demonstration mission is needed which demonstrates
  deployment of long tethers, stability and manoeuvreability of the
  spinning tether rig and use of E-sail for primary propulsion. The cost of
  the demonstration mission would be much less than that of the
  proposed Uranus mission.}

To probe the atmospheres of other giant planets and Titan, an
analogous mission architecture could be used. One could implement such
cluster mission by a fleet of identical probes designed for the worst
case heat load (Saturn, unless Jupiter is also included) and the
coldest cruise-phase thermal environment (Neptune). Alternatively, one
could optimise each probe to its specific mission, with some saving in
total mass and some increase in design cost.

\section{Acknowledgement}


The Electric Sail work is partly supported by the European Union
FP7/2010 grant 262733, the Academy of Finland (grant 250591) and the
Magnus Ehrnrooth Foundation.









\clearpage

\end{document}